\documentclass{article}
\usepackage{savesym}
\usepackage{amsmath}
\savesymbol{iint}
\usepackage[utf8]{inputenc}
\usepackage[numbers]{natbib}
\usepackage{graphicx}
\usepackage{amssymb}
\usepackage{authblk}
\usepackage{esint}
\usepackage{bbm}
\usepackage[margin=0.75in]{geometry}
\restoresymbol{TXF}{iint}

\title{The Geometry of a Quantum Circuit and its Impact on Electromagnetic Noise}
\author[1,2,*,$\dagger$]{Raina J. Olsen}
\author[2]{Mohammadreza Rezaee}
\author[1]{Radhakrishnan Balu}

\affil[1]{Computer and Information Sciences Directorate, Army Research Laboratory, Adelphi, MD, 21005-5069, USA}
\affil[2]{QSpice Labs Inc., \#2052, 105 Saint George St., Toronto, ON M5S 3E6, Canada}

\affil[$\dagger$]{Corresponding author rjo@qspicelabs.com}
\begin{document}

\maketitle

\begin{abstract}
Here we show that to quantize any lumped element circuit, the  circuit geometry must be included in a mathematical model of either the circuit fluxes or the circuit charges.  By geometry of the circuit, we refer to the so-called parasitic inductances and capacitances that arise from the details of the circuit layout, which are well known to create noise in classical circuits.  In contrast, the classical lumped element model describes only the topology of the circuit, which defines how different finite element variables are connected to one another by circuit components.  By geometry we also refer to the fact that the quantum variables define the circuit geometry - some are outside the wire,  some are inside the wire, and some are at boundary of the wire.  Just as with classical circuits, these effects create noise; this noise arises in the form of high frequency components in the Hamiltonian that are difficult to accurately simulate using a lumped element model.  The presentation is appropriate for undergraduate electrical and computer engineering students learning about quantum computing and physicists learning about electrical circuits.
\end{abstract}

\section{Introduction}\label{intro}

Humanity is currently nudging towards widespread industrialization of quantum computers.  As such, practitioners of quantum mechanics, a field which has long been treated as a high-level science, are being pushed towards a more engineering-oriented approach.  While quantum mechanics is essential to central concepts of electrical engineering, such as electrical conduction \cite{olsen2010} and semiconductor devices \cite{steele1957}, simplified versions of these concepts are usually taught to engineers.  For instance, the Drude model \cite{drude1900}, which was actually invented \emph{before} quantum mechanics, is still often used for describing electrical conduction.  Similarly, semiconductor devices have long been taught using band theory \cite{nussbaum1962} without ever showing how these bands arise from quantum interference \cite{olsen2010}.

Since quantum mechanics is fundamental to the operation of quantum computers, it seems highly probable that such oversimplification will not be so useful for training the new breed of electrical and computer engineers that will be designing and using superconducting quantum computers \cite{mermin,merminbook}.  Media is rife with excessively simple explanations of the basic principles of quantum computing, such as that a Qbit \cite{mermin,merminbook} can be both zero and one at the same time, so the computer does every possible computation at once.  These kinds of statements may be useful for journalists to hint at the power of quantum computing, but they are confusing and misleading for the students that should one day be designing these powerful computers.  For instance, while it may be true that an ideal quantum computer can be made to perform every possible computation at the same time, it is also true that only one result can ever be measured.  It is only through careful engineering of quantum interference (in the case of gate model architectures) or quantum tunneling (in the case of annealing architectures) that correct answers to difficult problems can be obtained more efficiently than with classical computers.

\begin{figure}[ht!]
  \centering
\includegraphics[scale=0.8]{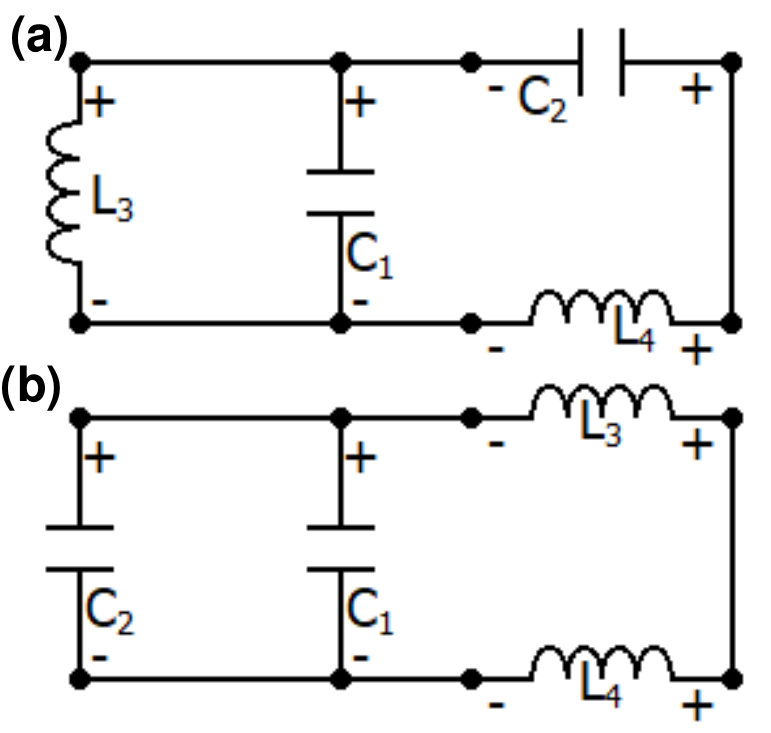}
\caption{Two lumped element circuits with the same topology.  Circuit (a) can be quantized using current methods \citet{devoret}, while circuit (b) cannot. }
\label{topo}
\end{figure}

The lumped element model has long been used as the most simple model of classical circuits.  A method of quantizing lumped element circuits described by \citet{devoret} has been used by the superconducting quantum computing field for decades.  Curiously, the relationship between the circuit topology and the component types determines whether or not a circuit can be quantized using this method.   For instance, the two circuits shown in Figure \ref{topo} have identical topologies; they differ only in the placement of the components within this topology.  But only the top one may be quantized.   And yet classically, both circuits have valid solutions.

It is an inarguable truth that \emph{reality is quantum}.  The correspondence principle says that classical descriptions are always an approximation of that quantum reality.  The fact that some lumped element circuits cannot be quantized using current methods means that there must be some fundamental piece of our quantum reality missing from the lumped element model that cannot be approximated away if we wish to describe the quantum mechanics of the circuit.  

Here, we show how to quantize lumped element circuits, demonstrating improvements to the method of \citet{devoret} that allow all lumped element circuits to be quantized, as long as the parasitic components that arise from the geometry of the circuit are always included.   Our method permits simple and automated quantization of every lumped element circuit.  An online quantization tool that uses our method can be accessed by students in order to explore and understand quantum circuits \footnote{The automated online quantization tool  can be found at this website: https://app.qspicelabs.com}.

Our finding that the circuit geometry must be included for a complete quantum description of circuits is an eminently intuitive result with deep physical meaning.  A circuit controls the flow of currents and strength of fields, but the fields arise from the properties of free space itself.  Thus it makes physical sense that complete description of the circuit must describe not only the topology - how the components are connected - but the geometrical arrangement of those connections in free space.  Practitioners of quantum circuits have long known through experience that the circuit layout is extremely important to the noise characteristics of the final circuits.  Our results provide a quantitative way to connect this intuition to the circuit design process.

Understanding the method described here requires a basic background in classical mechanics, specifically Lagrangian mechanics, in which the dynamics of a physical system are fully defined by the action of the system.  The constant of quantum mechanics, $\hbar$, is in units of action.  Any system may be quantized by defining its Lagrangian.  Application of this familiar technique of system quantization to the lumped element circuit model is a natural way to demonstrate principles of quantum mechanics by using a system very familiar to undergraduate electrical engineers.  Thus the work described here would work well as part of a course on quantum electrical and computer engineering.  We suggest the textbook ``Quantum Mechanics for Scientists and Engineers'' by Miller \cite{qeng}, the introduction to the Langriangian appropriate for undergraduate students of Hanc et al. \cite{hanc2004}, and the work by Mermin introducing quantum computing to computer scientists \cite{mermin,merminbook}.

We begin by reviewing background information, including lumped element circuits, the process of quantization, and the reason why some lumped element circuits cannot be quantized in Section \ref{back}.  Next we assess the assumptions of the lumped element model in Section \ref{assum} to identify those which might be revised to create the most simple, but complete model of quantum circuits.  In Section \ref{geom}, we show how geometric components can be added to the lumped element model to make it fully quantizable.  Finally, to validate the model we show the results of automated numerical simulations in Section \ref{results}.

\section{Background}\label{back}

Quantum circuits are made out of superconducting wires and components.  This superconductivity suppresses the natural dissipation of energy, and thus the dissipation of quantum information.  As such, the quantum circuit stays in the same quantum state, subject to time evolution under the rules of quantum mechanics, until an interaction changes that state.  That interaction may be intentional (through a control or measurement process), or an uncontrolled noisy interaction with the external environment.  The goal of this work is to show how a quantum circuit evolves under the rules of quantum mechanics between interaction events.

We need only consider capacitive and inductive components to discuss quantization of the circuit.  Thus in the present work, we neglect dissipative components (like resistors), sources, and non-linear components (like Josephson junctions), whose effects on quantum circuits have already been well-described by \citet{devoret}.  The issue of choosing a coordinate system, which requires the choice of ground node and a spanning tree, has also been well-described by \citet{devoret} and is discussed here only when necessary.

\subsection{Lumped element circuits} \label{classical}

In this Section, we review the traditional, classical methods of solving for the time evolution of lumped element circuits using coupled differential equations.

As a first approximation, circuits can simply be treated as a set of interconnected components using the lumped element model.  Each component has a type that determines its behavior -  inductors store energy in magnetic fields, while capacitors store energy in electric fields - and a number of terminals that can be connected to other components.  A circuit topology must also be defined; to be a single circuit,  wires must interconnect all the terminals in such a way that a path can be found between any two components.  Using this description of a circuit along with Kirchhoff's circuit laws, the behavior of the corresponding real circuit being modeled can often be accurately predicted and understood.

\begin{figure}[ht!]
  \centering
\includegraphics[scale=0.8]{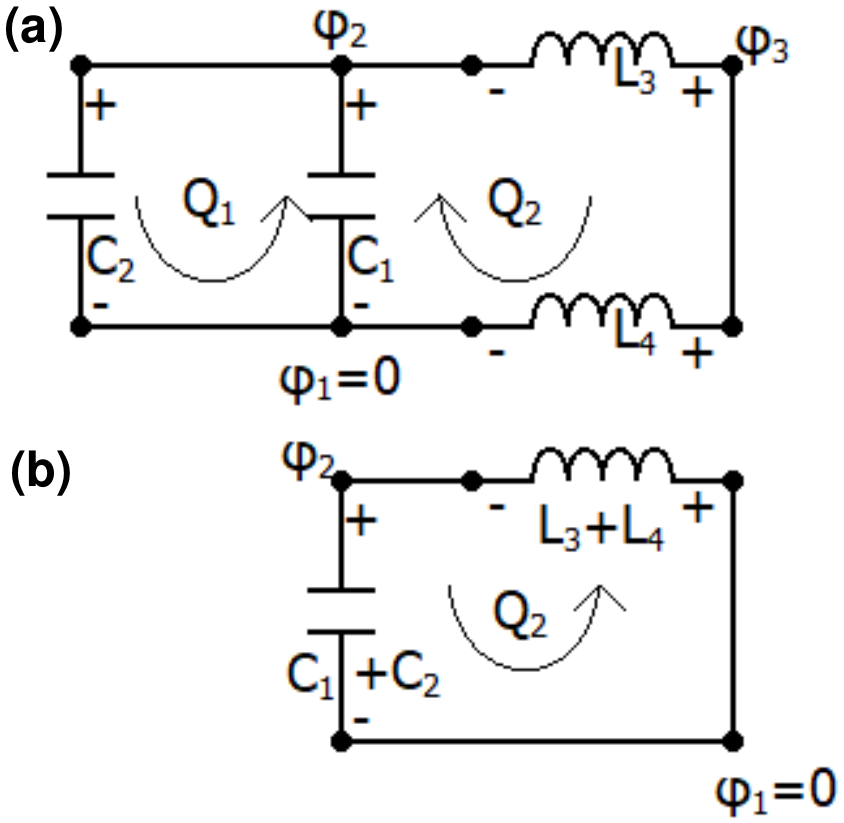}
\caption{A lumped element circuit that cannot be quantized in any representation using current methods. (a) The circuit.  The node flux and loop charge variables used in the solution are shown.  (b) The circuit reduced to a representation which can be quantized, but with loss of information about node 3 and loop 1.}
\label{unquan}
\end{figure}

These lumped element circuits are described by only a few degrees of freedom (DOF).  Consider the circuit shown in Figure \ref{unquan}(a), which is composed of four components, $c=4$.  In the classical lumped element model, either the voltages ($V$) across or the currents ($i$) through each component can be used as the degrees of freedom.  For each component type, different equations relate these two quantities: for a capacitor with capacitance $C$, $i=C\dot V$, and for an inductor with inductance $L$, $V=L \dot i$. 

But these component degrees of freedom are highly dependent on one another.  Independent DOF can be defined instead.  Each circuit has $n$ nodes and $l$ loops.  In the classical lumped element circuit model, the DOF are defined as the voltages $V_n$ at the $n$ circuit nodes  and the currents $i_l$ around the $l$ circuit loops.   One of the nodes is always defined as the ground node, which is at a constant voltage of zero, leaving $n-1$ node variables.  The number of loops is equal to $l=c+1-n$, for a total of $c$ independent degrees of freedom.  

Here we will find it more convenient to  modify the node voltages and loop currents slightly to node fluxes and loop charges by taking time integrals,
\begin{eqnarray}
\label{phidef}\phi_n&=& \int_{-\infty} ^t V_n(t') dt',\\
\label{Qdef}Q_l&=& \int_{-\infty} ^t i_l(t') dt.
\end{eqnarray}
In our four component circuit, there are two non-grounded nodes and two loops.  The node fluxes and loop charges are indicated on the diagram in Fig. \ref{unquan}(a).

Let us consider the classical solution of the lumped element circuit shown in Figure \ref{topo}(a).  Working in the node representation, we write equations using Kirchhoff's current law, which says that the currents arriving at each node sum to zero.    For our circuit, these equations at nodes 2 and 3 are,
\begin{subequations}
\begin{eqnarray}
\label{cn1}i_1+i_2-i_3=C_1 \ddot \phi_2 +C_2 \ddot \phi_2 + \frac{1}{L_3} (\phi_3-\phi_2) &=& 0,  \\
\label{cn2}i_4+i_3=\frac{1}{L_4} \phi_3 + \frac{1}{L_3} (\phi_3-\phi_2) &=& 0,
\end{eqnarray}
\end{subequations}
where $i_1,i_2\dots$ are the currents through each component.

To get the flux (or voltage) across a particular component, we have used differences between the flux at the nodes the component is attached to.  For instance, the inductor $L_3$ goes between nodes 3 and 2, thus the flux across the inductor is $\phi_3-\phi_2$.  Here are all the definitions for the circuit.
\begin{subequations}
\begin{eqnarray}
\label{fc1} \phi_{C1}&=& \phi_2  \\
\phi_{C2}&=& \phi_2  \\
\phi_{L3}&=& \phi_3-\phi_2 \\
\label{fl4} \phi_{L4}&=& \phi_3
\end{eqnarray}
\end{subequations}

Solving Eq. \ref{cn2} for $\phi_2$ and inserting into Eq. \ref{cn1}, we obtain the following equations,
\begin{eqnarray}
-(C_1 +C_2) \ddot \phi_2 &=& \frac{1}{L_3+L_4} \phi_2 ,  \\
\phi_3&=&\frac{L_4}{L_3+L_4} \phi_2,
\end{eqnarray}
for which the solution is,
\begin{subequations}
\begin{eqnarray}
\label{omega} \omega&=&\frac{1}{\sqrt{(C_1+C_2)(L_3+L_4)}}, \\
\label{sol} \phi_2&=&A \cos{(\omega t + \theta)}, \\
\phi_3&= &\frac{L_4}{L_3+L_4} A e^{\omega t + \theta},
\end{eqnarray}
\end{subequations}
where $A$ and $\theta$ are constants to be determined by the initial conditions.

Note that even though there are two independent node variables ($V_2,V_3$ or $\phi_2,\phi_3$), the solution has only one frequency and the solution for node 3 is simply a constant times the solution of node 2.  This happens because there are no capacitors attached to node 3,  only inductors. Thus there is no differential equation for $\phi_3$. Variables like this are often referred to as \emph{passive} variables \cite{devoret}.  We can remove the passive node by combining the series inductors and parallel capacitors in the circuit in Figure \ref{unquan}(a) to obtain the circuit shown in Figure \ref{unquan}(b).  This circuit has the same solution of Eq. \ref{sol}.

For comparison, we invite the reader to find the solution for the circuit shown in Figure \ref{topo}(a) instead.  For this circuit, one finds there are two frequencies - one for each independent variable - and that the solution for node 3 is a differential function of the solution for node 2.  None of the nodes are passive because there are capacitors attached to all nodes.

Classical lumped element circuits can also be solved using the loop representation.  For each loop, we  write equations using Kirchhoff's voltage law, which says that the voltages across each of the components that surround a loop sum to zero.    For our circuit, these equations are,
\begin{subequations}
\begin{eqnarray}
\label{lp1}-V_1+V_2&=& \frac{1}{C_1} (Q_1+Q_2) + \frac{1}{C_2} Q_1   = 0,  \\
\label{lp2}-V_1-V_3+V_4&=&  \frac{1}{C_1} (Q_1+Q_2) + L_3 \ddot Q^2 _2 + L_4 \ddot Q^2 _2    = 0,
\end{eqnarray}
\end{subequations}
We leave it is as an exercise for the reader to complete the solution in the loop representation, showing that the frequency $\omega$ and the solutions are identical to the solution in the node representation.

But we can immediately see within Eqs. \ref{lp1} and \ref{lp2} that there is no differential equation for $Q_1$.  This makes loop 1 a passive variable.  Loops are passive when they have no inductors around them.

\subsection{Lagrangian mechanics for lumped element circuits}

Next, we present an alternative method to solve for a lumped element circuit using Lagrangian mechanics.  To discuss the method, we will consider the simplest possible circuit, shown in Fig. \ref{unquan}(b), consisting of only one capacitor and one inductor connected in parallel.  The circuit has one loop and one non-grounded node.

The Lagrangian is defined  as the kinetic energy minus the potential energy,
\begin{eqnarray}
 \mathcal{L}=KE-PE
\end{eqnarray}
where the potential energy depends only on the variables themselves and the kinetic energy depends on the time derivative of the variables.    

Working in the node flux representation and using the classical definitions for the energy stored in a capacitor or inductor
\begin{eqnarray}
E_C&=&\frac{CV^2}{2}=\frac{Q^2}{2 C}, \\
E_L&=&\frac{\phi^2}{2 L}=\frac{LI^2}{2},
\end{eqnarray}
the Lagrangian for our simple circuit becomes,
\begin{eqnarray}
\label{lsimp} \mathcal{L}=\frac{(C_1+C_2) \dot{\phi}^2}{2}-\frac{ \phi^2}{2 (L_1+L_2)}.
\end{eqnarray}
The classical evolution of the system is determined by the Euler-Lagrange equation,
\begin{equation}
\frac{d}{dt} ( \frac{\partial  \mathcal{L}}{\partial \dot{\phi}} ) = \frac{\partial \mathcal{L}}{\partial \phi}.
\end{equation}
For a more complicated Lagrangian with many degrees of freedom, there will be a Euler-Lagrange equation for each variable. For our simple circuit, the Euler-Lagrange equation is
\begin{equation}
\ddot{\phi}=-\frac{1}{(C_1+C_2)(L_3+L_4)} \phi,
\end{equation}
which has same solution shown in Eq. \ref{sol}.  Thus we have shown that the Lagrangian method produces the same answer as the classical method  used in Section \ref{classical} for our circuit.

\subsection{Quantization of lumped element circuits}

The guiding principle of quantum circuits is that some macroscopic systems, those composed of effectively infinite numbers of particles, can still be accurately described by only a few degrees of freedom (DOF) evolving according to the rules of quantum mechanics.  It is not the reduction of a large system to a few DOF that makes the description quantum; we already reduced the system to a few DOF using the classical treatment in Section \ref{classical}.

What makes the circuit quantum is that these DOF can be expressed as pairs of canonically conjugate variables which must then be related to one another by the Heisenberg uncertainty principle.   The Heisenberg uncertainty principle relates the precision with which we can know the value of one variable (quantified by standard deviation $\delta \phi$) to the precision with which we can know the value of its conjugate variable by,
\begin{equation}
\label{heis} \delta \phi \delta \mathbbm{q} \ge \frac{\hbar}{2}.
\end{equation}

To be a conjugate pair, one variable of the pair must be the derivative of the action with respect to the other variable of the pair.  The action is defined using the Lagrangian as, 
\begin{equation}
S=\int_{t_2}^{t_2} \mathcal{L} dt ,
\end{equation}
and has dimensions of energy$\times$time. That is the reason why pairs of flux (dimensions of (energy$\times$time)/charge) and charge variables are used to quantize a circuit - because they are conjugate pairs.  In contrast, voltage and current do not form a conjugate pair.  

In this work, we consider only time independent Lagrangians. In this case, the conjugate variables can be obtained directly from the Lagrangian as,
\begin{equation}
\label{conjn} \mathbbm{q}_n=\frac{\partial \mathcal{L}}{\partial \dot{\phi_n}}.
\end{equation}

For the Lagrangian in Eq. \ref{lsimp}, we obtain the conjugate variable,
\begin{equation}
\mathbbm{q}=(C_1+C_2) \dot \phi_2
\end{equation}
From this definition, we can see that  the physical interpretation of the conjugate variable $\mathbbm{q}$ is the charge stored on the plates of all of the capacitors attached to that node.  As such, the $\mathbbm{q}_n$ variables are called capacitive node charges.  

The Hamiltonian, which gives the total energy of the system, is obtained from the Lagrangian through the Legendre transformation using the conjugate variables,
\begin{equation}
H= \sum_n \mathbbm{q}_n \dot{\phi_n} -\mathcal{L}.
\end{equation}

To quantize the system, one simply replaces each pair of conjugate variables $\phi,\mathbbm{q}$ with a pair of non-commuting operators $\hat{\phi},\hat{\mathbbm{q}}$,
\begin{equation}
 [\hat\phi, \hat{\mathbbm{q}}] = i \hbar.
\end{equation}
The Hamiltonian itself, $H$,  is also replaced with an operator $\hat{H}$.  The time evolution of the system can then be calculated using the Hamiltonian according to the rules of quantum mechanics.

\subsection{Unquantizable lumped element circuits} \label{quant}

Now let us return to the circuit shown in Figure \ref{unquan}(a) and see why it cannot be quantized.  The Lagrangian for this circuit is,
\begin{eqnarray}
\label{lag1n} \mathcal{L}=\frac{C_1 \dot{\phi_2}^2}{2}+\frac{C_2 \dot{\phi_2}^2}{2}-\frac{ (\phi_3-\phi_2)^2}{2 L_3}-\frac{ \phi_3^2}{2 L_4}.
\end{eqnarray}
We then obtain for the conjugate variables,
\begin{subequations}
\begin{eqnarray}
\label{q2} \mathbbm{q}_2&=& (C_1+C_2) \dot\phi_2,\\
\label{q3} \mathbbm{q}_3 &=& 0.
\end{eqnarray}
\end{subequations}
Now we can see the problem with this  particular circuit; the conjugate variable for node 2 is not actually a variable, it is simply always equal to zero.  Since we cannot identify a pair of canonical conjugate variables for node 2, the circuit cannot be quantized and we cannot determine the time evolution of this variable according to the rules of quantum mechanics.  

Let us discuss the physical meaning of this problem.  A quantum description of the circuit means that pairs of canonical conjugate variables are related to one another by the Heisenberg uncertainty principle (Eq. \ref{heis}).  If the conjugate variable is always precisely equal to zero, then its measurement error is also $\delta \mathbbm{q}_2=0$.  Inserting this in Eq. \ref{heis}, we obtain an infinite measurement error, $\delta \phi_2=\infty$, for the flux at node 2.  This means that the flux at node 2 is completely undetermined.  If we measure it, we might get anything.  This is unphysical nonsense, so there must be something we are missing.

What is the origin of this issue?  Recall that the physical interpretation of $\mathbbm{q}_n$ is the charge stored on the plates of all of the capacitors attached to that node.  Thus the reason that the conjugate variable at node 3 is undefined is because the node is passive - there are no capacitors attached to it.  

(One might think to avoid this problem by making node 2 the ground node.  We leave it is as an exercise for the reader to show why this solution does not work.)

Now let us attempt to instead quantize the circuit in the loop charge representation.  Using the derivative of Eq. \ref{Qdef}; $I_l=\dot Q_l$; the Lagrangian is
\begin{eqnarray}
\mathcal{L}=\frac{L_3 \dot{Q_2}^2}{2}+\frac{L_4 \dot{Q_2}^2}{2}-\frac{ (Q_1-Q_2)^2}{2 C_1}-\frac{ Q_1^2}{2 C_2}.
\end{eqnarray}
Note that in the node flux representation, the kinetic energy is the energy stored in the capacitors.  In contrast, here in the loop charge representation, the kinetic energy is the energy stored in the inductors.  The conjugate variables, 
\begin{equation}
\label{conjl} \mathcal{\Phi}_n=\frac{\partial \mathcal{L}}{\partial \dot{Q_n}}, 
\end{equation}
are called inductive loop fluxes.  For our circuit, they are
\begin{subequations}
\begin{eqnarray}
\label{p1} \mathcal{\Phi}_1&=& 0,\\
\label{p2} \mathcal{\Phi}_2 &=& (L_3+L_4) \dot Q_2.
\end{eqnarray}
\end{subequations}
In this representation, we find that the conjugate variable for loop 1 is undefined.   Similarly, this is because loop 1 is passive - there are no inductors around the loop.  Since no conjugate variable is defined for the charge around loop 1, we cannot determine the time evolution of this variable according to the rules of quantum mechanics.

If we want to solve for the properties of this particular circuit, we can reduce it to the circuit in Figure \ref{unquan}(b) by combining the two capacitors and the two inductors.  This reduced circuit has a capacitor connected to every node and so is quantizable in the node flux representation, as we showed in the last Section.  Similarly, the reduced circuit has an inductor around every loop, so is quantizable in the loop charge representation.  

\begin{figure}[ht!]
  \centering
\includegraphics[scale=0.8]{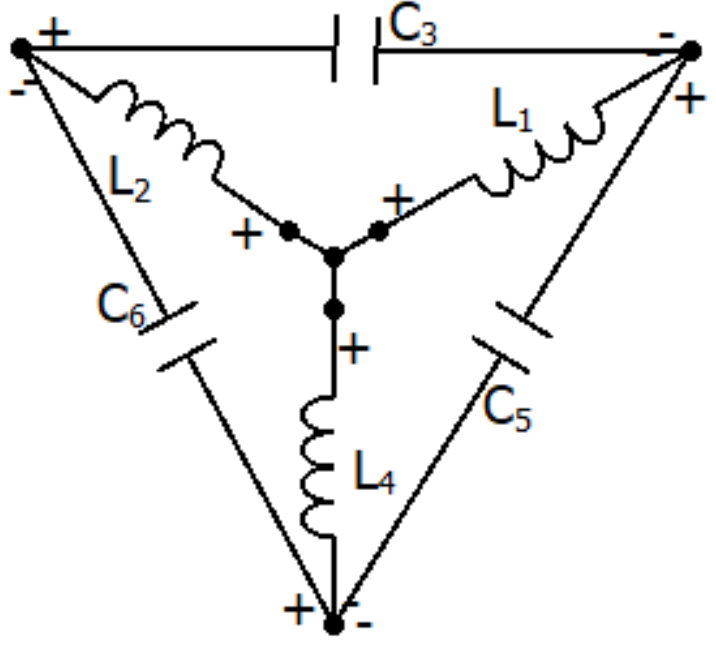}
\caption{Irreducible lumped element circuit that cannot be quantized. }
\label{unquan2}
\end{figure}

However, by reducing the circuit in this way, we lose information about one of the degrees of freedom in our solution.  And not every unquantizable circuit can be reduced to a circuit which is solvable.  Figure \ref{unquan} shows another circuit which cannot be quantized.  It cannot be quantized in the node flux representation because the central node does not have a capacitor connected to it.  Likewise, it cannot be quantized in the loop charge representation because the outermost loop, the one that runs around the outside edge of the circuit, does not have an inductor around it.  And, this circuit cannot be reduced to a simpler circuit which is solvable.

As such, we cannot count on being able to deal with the problem of unquantizable circuits by reducing the circuit to a simpler circuit that is solvable.  However, we have chosen a circuit that \emph{is} can be reduced so that we have a way to verify the method we will present in the rest of this manuscript, which allows us to quantize any lumped element circuit whose geometry is defined.

To summarize the problem, using the present method for quantizing circuits, we are unable to quantize circuits in the node flux representation when there are passive nodes which are not attached to any capacitor.  Similarly, we are unable to quantize circuits in the loop charge representation when there are passive loops which do not go through an inductor.  Circuits which contain both passive nodes and passive loops cannot be quantized in any representation using this method, even though we can write classical equations determining the time evolution of the circuit.  Quantum mechanics is a more fundamental theory than classical mechanics.  If we can describe the system classically, there must be a way to describe it quantum mechanically.

\section{Discarding assumptions of the lumped element model} \label{assum}

In the classical world of circuits, it is already well known that the lumped element  is nothing more than an approximation.  To better understand why we are having issues quantizing some lumped element circuits, let us review the three assumptions of the lumped element model to see which may need to be discarded in order to quantize all lumped element circuits.

The first assumption of the lumped element model is that the length scale of the circuit $r$ is smaller than the wavelength of the radiation,
\begin{equation}
r \ll \lambda.
\end{equation}
In typical quantum circuits, the size of the circuit elements are on the micrometer scale.  Inductances are at the nano-Henry scale, while capacitances are at the pico-Farad scale.  This leads to GHz scale frequencies; $f=\omega/2\pi=1/2\pi\sqrt{LC}=5$ GHz.  These microwaves have wavelengths in the centimeter range, which is indeed orders of magnitude larger than the length scale of the circuit.  Thus there does not appear to be an issue with this assumption.

The second assumption of the lumped element model is that outside of the conducting wires, the magnetic flux $\Phi_B=\vec{B} \cdot \vec{A}$ can be taken to be constant,
\begin{equation}
\label{assumf} \frac{\partial \Phi_B }{\partial t} \simeq 0.
\end{equation}
This assumption is normallly used to obtain Kirchhoff's voltage law.  Here we will show the derivation of a modified form of Kirchhoff's voltage law by taking this term to be small, but not necessarily zero. The derivation starts with the integral form of the Maxwell-Faraday equation,
\begin{equation}
\label{mxf}\oint_{\partial S} \vec{E} \cdot d\vec{r} = \iint_S \frac{\partial \vec{B} }{\partial t} d\vec{A} =  \frac{ \partial {\Phi}_l}{\partial t}.
\end{equation}
where $S$ is a surface bounded by the contour $\partial S$, with the contour $\partial S$ chosen to run around a loop in the circuit.    Thus ${\Phi}_l(t)$ is the magnetic flux through the loop. (Note that unlike the inductive loop flux $\mathcal{\Phi}$ defined in Eq. \ref{conjl}, which is the flux in the inductors that the loop of wire passes through, ${\Phi}_l(t)$ is the magnetic flux going through the surface $S$.)

Since the contour $\partial S$ goes around a loop in the circuit, it must pass through at least two components.  The voltage across each component around the loop can be defined as,
\begin{equation}
V_{c}=\int_{c\in \partial S}\vec{E} \cdot d\vec{r} .
\end{equation}
Substituting this into Eq. \ref{mxf}, we obtain Kirchhoff's voltage law,
\begin{equation}
\label{kvl} \sum_{c\  around \ l} V_{c} =  \frac{ \partial {\Phi}_l}{\partial t} \simeq 0,
\end{equation}
where the r.h.s. is normally taken to be identically zero in the lumped element model.  As we explained in Section \ref{classical}, we prefer to work with charge and flux variables rather than current and voltage variables because they form conjugate pairs.  We can integrate both sides of Eq. \ref{kvl} to obtain the following `flux law'.
\begin{equation}
\label{mkvl} \sum_{c\  around \ l} \phi_{c} = {\Phi}_l.
\end{equation}
The lumped element model (specifically Kirchhoff's voltage law) assumes that ${\Phi}_l$ is a constant.  If we discard this assumption, then ${\Phi}_l$ becomes a time varying value.

The third assumption of the lumped element model is that inside the conducting wires, the charge $q$ is constant,
\begin{equation}
\label{assumq} \frac{\partial q }{\partial t}\simeq 0
\end{equation}
This assumption is normally used to obtain Kirchhoff's current law.  Here we will show the derivation of a modified form of Kirchhoff's current law by taking this term to be small, but not necessarily zero. The derivation starts  with the charge continuity equation (which is itself derived from Ampere's law),
\begin{equation}
\label{cont} \nabla \cdot \vec{J} =\frac{\partial \rho }{\partial t},
\end{equation}
where $\rho$ is the charge density and $J$ is the current density.  Let us choose a small volume $V$ that completely encloses a node in the circuit (but does not enclose any part of any component) with bounding surface $\partial V$.
\begin{equation}
\label{vol2} \iiint_V \nabla \cdot   \vec{J} =\iiint_V \frac{\partial \rho }{\partial t}
\end{equation}
Now integrate the right hand side of Eq. \ref{vol2} over that volume, and apply the divergence theorem to the left hand side, obtaining
\begin{eqnarray}
\label{cont2} \oiint_{\partial V}  \vec{J} \cdot d\vec{A} =\frac{\partial q_n(t). }{\partial t},
\end{eqnarray}
where $q_n(t)$ is the charge at the node.  (Note that unlike the capacitive node charge $\mathbbm{q}$ defined in Eq. \ref{conjn}, which is the charge on the capacitors attached to the node, $q_n(t)$ is the charge collected at the node itself.) Since the surface contour $\partial V$ goes around a node in the circuit, it must pass through at least two conductors that go into the components that are attached to the node.  The current going into each component can be defined by,
\begin{equation}
i_{c}=\int_{c\in \partial V} \vec{J} \cdot d\vec{A} .
\end{equation}
Substituting this into Eq. \ref{cont2}, we obtain the usual form of Kirchhoff's current law,
\begin{equation}
\label{kcl} \sum_{c \ attached \ n} i_{c} = \frac{\partial q_n }{\partial t} \simeq 0,
\end{equation}
where the r.h.s. is normally taken to be identically zero in the lumped element model.  As we described in the last Section, we prefer to work with charge and flux variables rather than current and voltage variables because they form conjugate pairs.  We can integrate both sides of Eq. \ref{kcl} to obtain the following `charge law'.
\begin{equation}
\label{mkcl} \sum_{c \ attached \ n} Q_{c} = q_n .
\end{equation}
The lumped element model (specifically Kirchhoff's current law)  assumes that $q_n$ is a constant.  If we discard this assumption, then $q_n$ becomes a time varying value.

In this section, we have discussed the second and third assumptions of the lumped element model, which are used in the classical lumped element model to derive Kirchhoff's voltage and current laws.  In the quantum lumped element model, where we deal with conjugate flux and charge variables instead of voltage and current variables, these assumptions amount to taking the flux through a loop and the charge at a node to be constants.  If we discard these assumptions, then the flux through a loop and the charge at a node must be treated as variables, even though they are very small.  In the next Section, we show how they can be included in the Lagrangian.

\section{Geometric components} \label{geom}

Now that we have reviewed the assumptions of the lumped element model, let us return to our unquantizable circuit.  In Section \ref{quant}, we were unable to quantize a circuit containing a loop which does not have any inductive component around it.  However, \emph{any} loop of wire has a self-inductance associated with it, which is a function only of the geometry of the loop.   This is often referred to as a `parasitic' inductance, terminology that refers to the assumption that these inductances are so small that they are normally neglected in the lumped element model.  Indeed, loops with sizes on the order of micrometers will have self-inductances at the pico-Henry scale, many orders of magnitude smaller than the components in these circuits which are designed to be inductive.  

But let us consider the fact that an inductor is a component specifically designed to enhance the natural ability of free space to store energy in a magnetic field in response to a changing electrical current.  As such, it is not so much that the self-inductances are particularly small; rather, it is that the inductances of the components designed to be inductive are particularly large.

Similarly, in Section \ref{quant}, we were unable to quantize a circuit containing a node which does not have any capacitor attached to it.  However, \emph{any} two electrical conductors have a capacitance between them, which is a function only of the relative geometry of the conductors.  However, these are normally so small that they can safely be neglected.  Nodes on the opposite sides of a circuit loop with sizes on the order of micrometers will have a capacitance between them of only femto or even atto-Farads.

Since both of these so-called `parasitic' components depend only on the geometry of the circuit, we refer to them here as `geometric components' instead.  To include these geometric components in the description of the circuit, we must do two things.  The first thing is quite straightforward; we must include the energy stored in these geometric components in the Lagrangian.  For the flux representation, after adding the geometric components to Lagrangian in Eq. \ref{lag1n}, we obtain
\begin{eqnarray}
\nonumber \mathcal{L}_n&=&\frac{C_1 \dot{\phi_2}^2}{2}+\frac{C_2 \dot{\phi_2}^2}{2} + \frac{C_{g32} (\dot{\phi_3}-\dot{\phi_2})^2}{2} + \frac{C_{g21} \dot{\phi_2}^2}{2} +  \frac{C_{g31} \dot{\phi_3}^2}{2}- \\
&&\frac{ (\phi_3-\phi_2)^2}{2 L_3}-\frac{ \phi_3^2}{2 L_4} - \frac{ \Phi_1^2}{2 L_{g1}} - \frac{ {\Phi}_2^2}{2 L_{g2}}.
\end{eqnarray}
Now let us check what the conjugate variables, $ \mathbbm{q}_n=\partial \mathcal{L} / \partial \dot{\phi_n}$, have become.
\begin{subequations}
\begin{eqnarray}
 \mathbbm{q}_2&=& (C_1+C_2+C_{g21}+C_{g32}) \dot\phi_2 - C_{g32} \dot\phi_3,\\
 \mathbbm{q}_3 &=&  (C_{g32}+C_{g31}) \dot\phi_3 - C_{g32} \dot\phi_2, \\
 \mathbbm{Q}_1 &=& \frac{\mathcal{L} }{\partial \dot{{\Phi}}_1}=0, \\
 \mathbbm{Q}_2 &=& \frac{\mathcal{L} }{\partial \dot{{\Phi}}_2}=0.
\end{eqnarray}
\end{subequations}
Now each node variable has a conjugate variable defined, and so they may be quantized.  But we have added two new variables, which give the flux through the loops.  These do not have conjugate variables defined.  So we appear to have simply pushed the problem back a stage.

One way that we could deal with this problem is to not treat the loop fluxes as variables, but instead treat them as constants, as other authors do \citet{devoret}.  This is justifiable as they are very small - indeed they are usually treated as constants to derive Kirchhoff's voltage and current laws.  Below in Section \ref{results} we show simulations that use this method.  

However, it is also informative to see what we can learn if we treat them as variables.  We have not yet explicitly considered Kirchhoff's laws, either here or in Section \ref{back} where we first wrote the Lagrangian for the circuit.  Recall the flux law, Eq. \ref{mkvl}, which we derived in Section \ref{assum} from Maxwell's equations
\begin{eqnarray}
\nonumber \sum_{c\  around \ l} \phi_{c} = {\Phi}_l.
\end{eqnarray}

Though we did not explicitly explain this, in Section \ref{back} we simply took ${\Phi}_l(t)$ to always be zero (making the same assumption used to derive Kirchhoff's voltage law) by defining the flux across each component in Eqs. \ref{fc1}-\ref{fl4} to be the difference between the flux at the two nodes on either side of the component.  For instance, consider the second loop, which goes through components $C_1$, $L_3$, and $L_4$.  Using Eqs. \ref{fc1}-\ref{fl4}, we obtain, 
\begin{eqnarray}
\nonumber \sum_{c\  around \ l=2} \phi_{c} &=& \phi_{C1} + \phi_{L3} - \phi_{L4} \\
\label{kir2}&=& (\phi_2)+(\phi_3-\phi_2)-(\phi_3) = 0
\end{eqnarray}
But by including geometric inductances, the flux through the loops have become non-zero, time varying quantities.  Eqs. \ref{fc1}-\ref{fl4} must be redefined to include these new variables using our flux law.  Here, we will use the following definitions,
\begin{subequations}
\begin{eqnarray}
\phi_{C1}&=& \phi_2 \\
\phi_{C2}&=& \phi_2 -{\Phi}_1 \\
\label{fluxl3} \phi_{L3}&=& \phi_3-\phi_2 \\
\label{fluxl4} \phi_{L4}&=& \phi_3 +{\Phi}_2
\end{eqnarray}
\end{subequations}
With these new definitions, Eq. \ref{kir2} becomes,
\begin{eqnarray}
\nonumber \sum_{c\  around \ l=2} \phi_{c} = (\phi_2)+(\phi_3-\phi_2)-(\phi_3+{\Phi}_2) = -{\Phi}_2,
\end{eqnarray}
thus fulfilling Kirchhoff's flux law (with the negative sign reflecting the fact that ${\Phi}_2$ was been defined in Figure \ref{unquan}(a) with a left handed circulation.)  The reader is invited to check that the flux law is also fulfilled for the first loop in the circuit.  \citet{devoret} gives a more in depth discussion of how these definitions can be determined for any arbitrary circuit by choosing a spanning tree. 

Our Lagrangian becomes, 
\begin{eqnarray}
\nonumber \mathcal{L}_n&=&\frac{C_1 \dot{\phi_2}^2}{2}+\frac{C_2 (\dot \phi_2 -\dot{\Phi}_1)^2}{2} + \frac{C_{g32} (\dot{\phi_3}-\dot{\phi_2})^2}{2} + \frac{C_{g21} \dot{\phi_2}^2}{2} +  \frac{C_{g31}(\dot \phi_3 +\dot{\Phi}_2)^2}{2}- \\
&&\frac{ (\phi_3-\phi_2)^2}{2 L_3}-\frac{ (\phi_3 +{\Phi}_2)^2}{2 L_4} - \frac{ \Phi_1^2}{2 L_{g1}} - \frac{ {\Phi}_2^2}{2 L_{g2}},
\end{eqnarray}
and the conjugate variables are now all defined,
\begin{subequations}
\begin{eqnarray}
 \mathbbm{q}_2&=& (C_1+C_2+C_{g21}+C_{g32}) \dot\phi_2 - C_{g32} \dot\phi_3  -C_2 \dot{\Phi}_1,\\
 \mathbbm{q}_3 &=&  (C_{g32}+C_{g31}) \dot\phi_3 - C_{g32} \dot\phi_2- C_{g31} \dot\Phi_2, \\
 \mathbbm{Q}_1 &=& C_2 \dot{\Phi}_1 - C_2 \dot\phi_2 ,\\
 \mathbbm{Q}_2 &=&  C_{g31} \dot{\Phi}_2 + C_{g31} \dot\phi_3.
\end{eqnarray}
\end{subequations}
Now that the conjugate variables have been defined, the circuit can be quantized.

Note that we have not included here the effect of parasitic mutual inductances between different loops in the circuit, a topic which the reader is invited to explore by defining an inductance matrix.   The reader may also be interested to find the solution for the circuit in the loop charge representation by including the geometric components in the Lagrangian. (Hint: a condition defining the net charge of the circuit must be used to solve for the system in the loop charge representation.)

\section{Results} \label{results}

Here we show solutions for both circuits of Figure \ref{topo} using the following values for the components: $C_1$=2 pico-Farad (pF), $C_2$=4 pF, $L_3$=1 nano-Henry (nH), $L_4$=3 nH.  The initial conditions are 2 mili-Volts (mV) across each capacitor in the circuit and 0 nano-Amperes (nA) through each inductor in the circuit.

In the last Section, we showed that any circuit could be fully quantized by treating the loop fluxes (node charges) as variables in addition to the node fluxes (loop charges).  But we also showed that taking the loop fluxes (node charges) as constants is equivalent to using Kirchhoff's voltage and current laws.  In the simulations shown here, we will use this latter method.  This topic will be discussed further in the next Section.

\begin{table}
\caption{Solution parameters for the circuit shown in Fig. \ref{topo}(a), which has no passive variables, in both the node flux and loop charge representations.}
\begin{center}
 \begin{tabular}{|c|c c c c||} 
 \hline
 representation&freq. (GHz) & variable & freq. (GHz) & variable \\ [0.5ex] 
 \hline\hline
 node & 3.56 & $\phi_2$ & 2.51 & $\phi_3$  \\ 
 loop & 2.51 & $Q_1$ & 3.56 &  $Q_2$ \\ 
 \hline
\end{tabular}
\end{center}
\label{table1}
\end{table}

\begin{figure}[ht!]
  \centering
\includegraphics[scale=0.8]{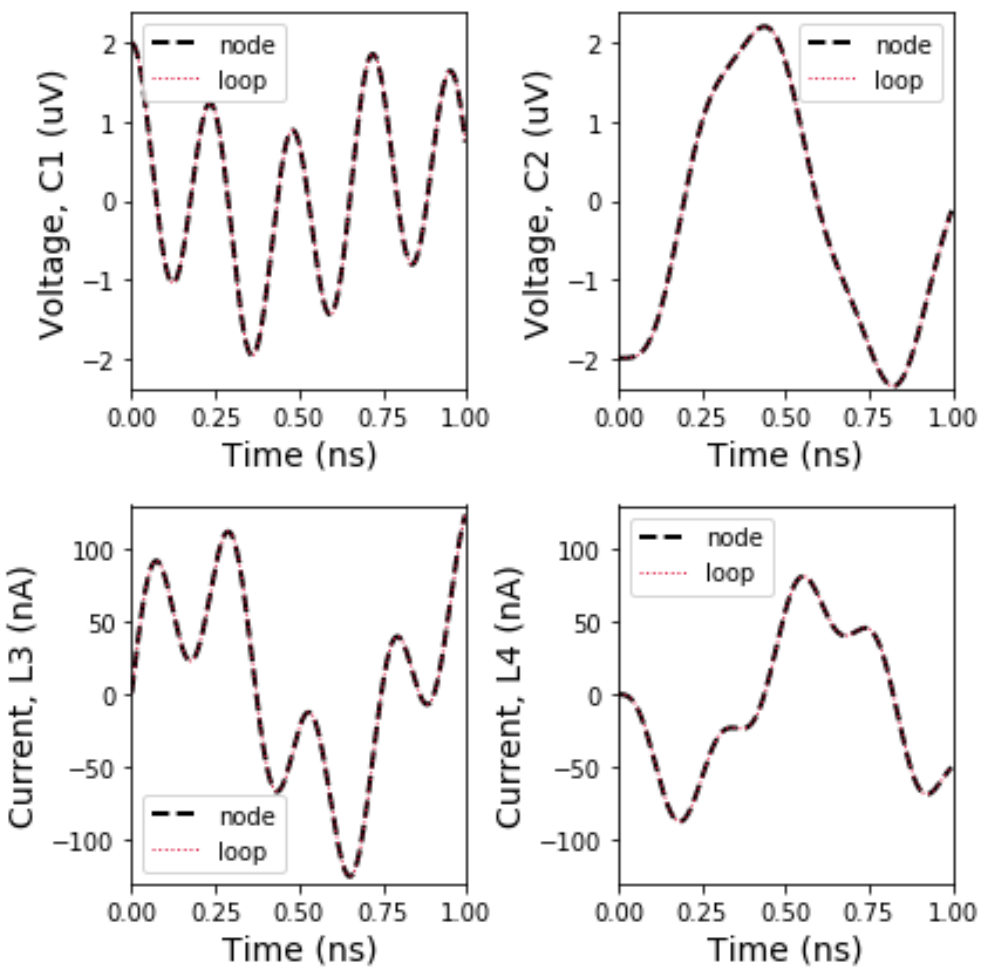}
\caption{Quantum simulations of the circuit shown in Fig. \ref{topo}(a), which has no passive variables. Solutions were calculated by time evolving the Hamiltonian calculated in both the node flux and loop charge representations.}
\label{qres}
\end{figure}

Figure \ref{qres} shows the results of simulations for the lumped element circuit shown in Fig. \ref{topo}(a) and Table \ref{table1} shows the parameters of the solution.  This is the circuit with no passive nodes or loops - there is a capacitor attached to each node and an inductor around each loop.  We invite the reader to derive the Hamiltonian for this circuit and compare it to our solution.    Table \ref{table1} shows that the solution describes oscillators with frequencies of 2.51 and 3.56 GHz.  These frequencies do not depend on whether the node flux or loop current representation is used.

However, these two oscillators are strongly coupled together.  As such, the solutions shown in Figure \ref{qres} are not just simple oscillators.  The voltage across $C_2$ has a complex shape with a frequency of approximately 1 GHz, which is the difference between the frequency of the two oscillators.  The voltage across $C_1$ also has a complex shape, with a frequency larger than the frequency of either oscillator.

\begin{figure}[ht!]
  \centering
\includegraphics[scale=0.8]{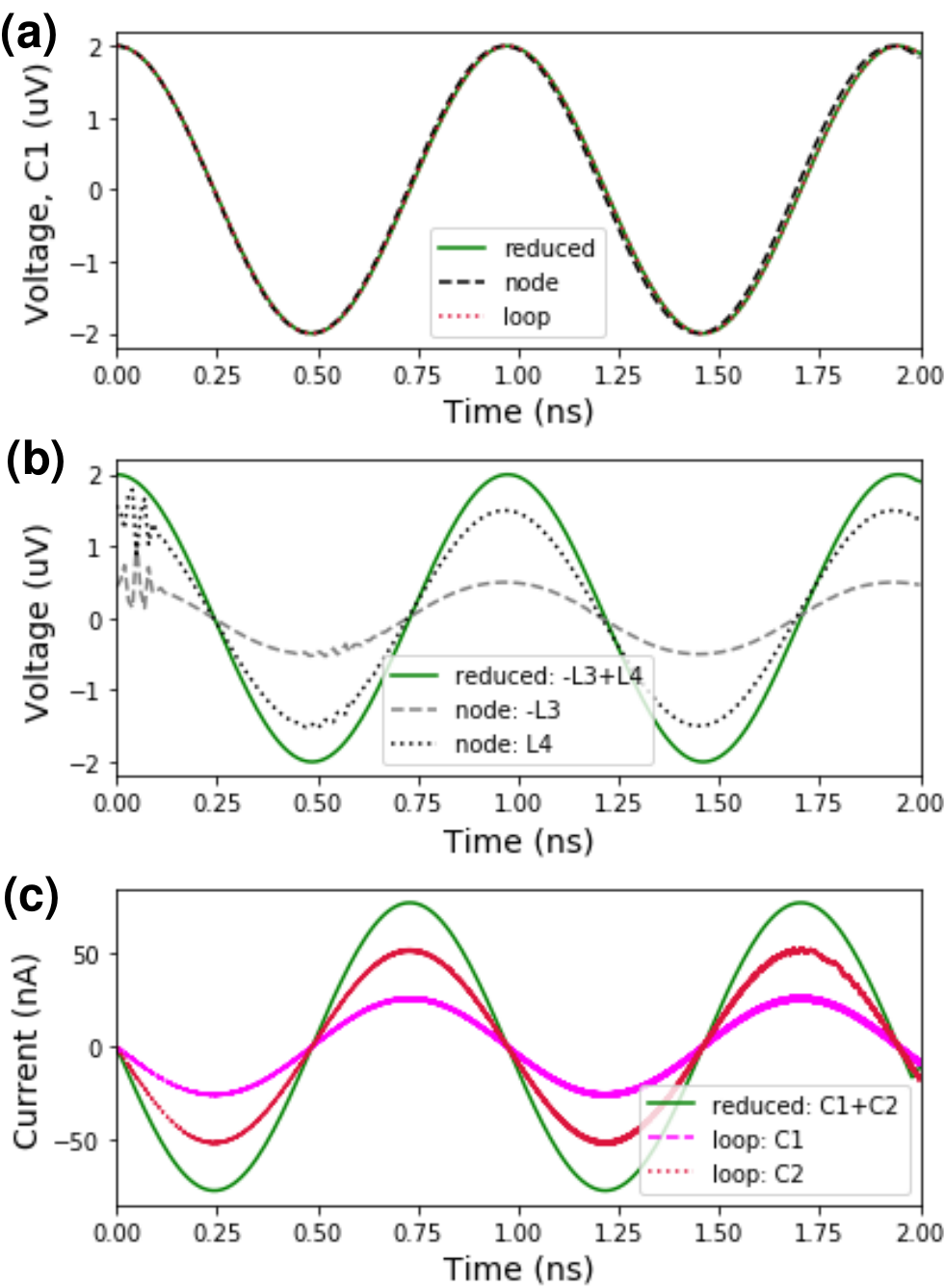}
\caption{Quantum simulations of the circuit shown in Fig. \ref{topo}(b), which contains passive variables. Simulations of the full circuit in both the node flux and loop charge representations are compared to the solution for the `reduced' circuit shown in Fig. \ref{unquan}(b).}
\label{unqres}
\end{figure}

Figure \ref{unqres} shows the results of simulations for the lumped element circuit shown in Fig. \ref{topo}(b) and Table \ref{table2} shows the parameters of the solution.  This is the circuit that has a passive node and a passive loop, and so cannot be quantized unless geometric components are included in the solution.  

For comparison, a simulation for the `reduced' circuit shown in Fig. \ref{unquan}(b) is also shown.  This reduced version of the circuit has only one capacitor with capacitance 6 pF - comprising the two parallel capacitors in Fig. \ref{unquan}(a).  Likewise, there is only one inductor with inductance 4 nH - comprising the two series inductors in Fig. \ref{unquan}(a).  Referring to Eq. \ref{omega}, we find that this produces a frequency of 6.46 rad/ns or 1.03 GHz.  The circuit solution plotted in Figure \ref{qres} does indeed show a simple oscillator with a frequency of 1.03 GHz.  However, the solution for this `reduced' circuit lost information about the individual components that were combined to create the simplified circuit.

\begin{table}
\caption{Solution parameters for the circuit shown in Fig. \ref{topo}(b), which contains passive variables.  Solutions of the full circuit in both the node flux and loop charge representations are compared to the solution for the `reduced' circuit shown in Fig. \ref{unquan}(b).}
\begin{center}
 \begin{tabular}{|c|c c c c||} 
 \hline
 representation&freq. (GHz) & variable & freq. (GHz) & variable \\ [0.5ex] 
 \hline\hline
  node, reduced & 1.03 & $\phi_2$ &  &  \\ 
 node & 2.05 & $\phi_2$ & 18221 & $\phi_3$  \\ 
 loop & 1.80 & $Q_1$ & 1378 &  $Q_2$ \\ 
 \hline
\end{tabular}
\end{center}
\label{table2}
\end{table}

Figure \ref{unqres} also shows solutions calculated using the methods presented here to quantize the full circuit.  A geometric capacitance of $C_{g23}$=8.9$\times 10^{-8}$ pF and a geometric self inductance of $L_{g1}$=1.0$\times 10^{-6}$ nH was used in these solutions.  These are reasonable values for nodes and loops with length scales on the order of a hundred micrometers composed of wires several micrometers wide.  Table \ref{table2} shows that the solutions for the full circuit have oscillator frequencies very different from the frequency of the reduced circuit.  Nevertheless, with the correct coupling between these different oscillators, the solutions for the voltage across $C_1$ shown in Figure \ref{unqres}(a) are nearly identical.

However, the inclusion of the geometric components does have an important impact on the solution for a circuit with passive nodes or loops.  Figure \ref{unqres}(b) shows the voltage across each of the inductors in the circuit.  The net voltage across  both inductors is a simple oscillator with a frequency of approximately 1 GHz.  But the voltage across each individual inductor shows high frequency oscillations.  These originate from tiny fluctuations in the charge of node 3.  Since the capacitance attached to this node is very small and $V=Q/C$, these tiny charge fluctuations cause visible high frequency noise in the voltage.  

Similarly, Figure \ref{unqres}(c) shows the current through each of the capacitors in the circuit.  The net current through both capacitors is a simple oscillator with a frequency of approximately 1 GHz.  But the current through each individual capacitor shows high frequency oscillations.  These originate from tiny fluctuations in the flux through loop 1.  Since the self inductance attached to this node is tiny and $I=\phi/L$, these small flux fluctuations cause visible high frequency noise in the current.

Here it is important to point out that the simulations in different representations \emph{do not produce the same result} for the circuit with passive nodes and loops.  The large-scale behavior is the same, but the high frequency noise depends not only on the representation (node or loop), but also on the spanning tree used (though we have not shown simulations with different trees here - the reader is invited to test some themselves.)  When the node representation is used, high frequency noise appears only in the voltage at node 3.  When the loop representation is used, high frequency noise appears only in the current around loop 1.

\section{Discussion} \label{summary}

In this paper we have demonstrated something that everyone who works with quantum circuits already knows through experience - that the geometry of the circuit can have a significant impact on noise in the circuit such that it is difficult to accurately predict using simulation exactly how the real circuit will respond.  This occurs through the same mechanisms that create noise in classical circuits - the parasitic components that arise from the circuit layout and the approximations contained within Kirchhoff's laws.

However, for students who come from a background in classical circuits and are just learning quantum mechanics, it is quite informative to see how this noise appears in the mathematics of the Hamiltonian.  Because the values of the geometric components tend to be orders of magnitude smaller than the lumped element components and because frequency is inversely proportional to these small component values, the frequency of passive nodes and loops tends to be much larger than the frequencies of the active nodes and loops.  A similar result is found if the full solution is considered, with loop fluxes (node charges) quantized rather than treated as constants.  The loop fluxes (node charges) also have much larger frequencies than the node fluxes (loop charges).

High frequency corresponds to high energy.  For instance, the frequencies of the passive variables in Table \ref{table2} are in the THz regime, corresponding to temperatures from dozens to hundreds of Kelvin.  In contrast, quantum circuits are generally operated at sub-Kelvin temperatures.  Thus the population of these high energy states will generally be quite small.  

\begin{figure}[ht!]
  \centering
\includegraphics[scale=1.0]{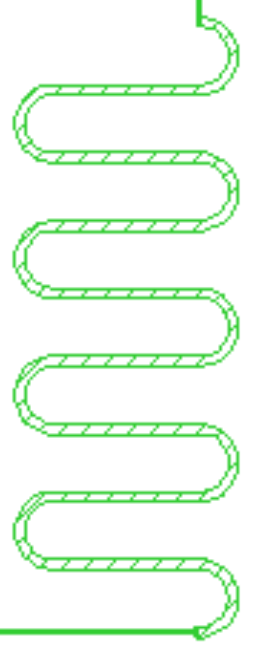}
\caption{Layout of a superconducting inductor drawn with KLayout.}
\label{real}
\end{figure} 

However, it is unphysical that these frequencies are really quite so high as we have found here.  Consider the realistic `S'-shaped inductor layout shown in Figure \ref{real}.  Instead of treating this inductor as a single lumped element, it would be more accurate to treat it as many inductors in series.  Nodes in the middle of this series of inductors are passive, unless we consider the capacitances between them.  Each curve of the `S' does have a small capacitance with the curve above and below it.  This is modeled as a small capacitor in parallel with each inductor.  Though these capacitors are small, they are not quite so small as the capacitor that goes between the node at the very top of the inductor and the node at the very bottom of the inductor.  Thus high frequencies will still appear in the solution, but they will not be at quite such high frequencies as the THz scale.

\begin{figure}[ht!]
  \centering
\includegraphics[scale=0.8]{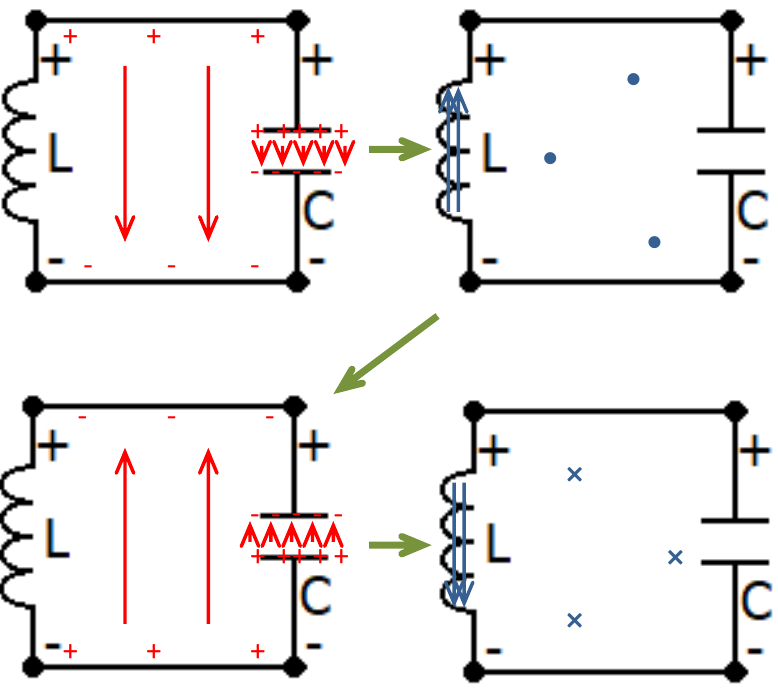}
\caption{Depiction of the electric and magnetic fields in and around an LC circuit during a single period of the oscillation between the lumped element capacitor and the lumped element inductor.}
\label{physical}
\end{figure}

What is the physical source of these high frequency components?  Consider the depiction of the electric and magnetic fields of an LC circuit shown in Figure \ref{physical}.  When there is a current in the circuit, as depicted on the right side of the Figure, then energy is stored within a strong magnetic field inside the inductor.  A current in a wire loop always creates a magnetic field inside that loop.  An inductor is constructed with many overlapping loops in order to strongly concentrate the magnetic field inside of it.  But no matter how perfect the inductor is made, the magnetic field can never be entirely confined within the inductor coil.  Magnetic flux will always appear within the loops of the circuit itself.  

Moreover, this magnetic flux within the circuit loop cannot simply be considered to be leakage from the inductor.  The inductor coil depicted in the Figure creates a magnetic field that lies parallel to the page, while the circuit loop creates a magnetic field that lies perpendicular to the page.

When energy is stored within a strong electric field inside the capacitor, as depicted on the left side of the Figure, this means that there is a voltage between the two nodes of the circuit.  Voltage is defined as the line integral of the electric field between the two nodes.  If the path of that line integral goes through the wires and through the capacitor itself, then the voltage is acquired through a strong electric field integrated over the short path between the two plates of the capacitor. But if the path goes through the center of the circuit loop instead, then the voltage will only be nonzero if the electric field outside of the capacitor is not zero.  The voltage is instead acquired through a weak electric field integrated over the path between the nodes, which is much longer than the distance between the capacitor plates.  No matter how perfect the capacitor is made, the electric field can never be entirely confined within its plates.  There will always be small electric fields between the nodes of the circuit.

At first thought, this stray electric field could simply be considered leakage from the capacitor.  However, our solutions show that these stray fields oscillate with a frequency much larger than the frequency of the design components and thus with a frequency much larger than the variation of a leakage field from the capacitor.  Stray electric fields could also arise from the collection of small charges at the nodes.  Normally it is assumed in the derivation of Kirchhoff's laws that the net charge inside a conductor is always exactly zero, and so these node charges must be zero.  But we discarded this assumption when deriving Eq. \ref{kcl}.  

The reasoning behind the assumption that node charges must be zero is that the electrons repel one another and so tend to end up as far away from one another as they can get, such as on the plates of the lumped element capacitors.  But this reasoning only applies at steady state.  The electrons move so fast that this steady state is usually achieved very quickly, much more quickly than the period of the oscillatory motions between the designed components.  But it is precisely in the case of quick changes between different steady states that high frequency components come into play when a Fourier expansion of the time changing field is considered.  

Therefore, small electric and magnetic fields appear outside the components, within the loops of the circuit itself.  These are not just stray fields that leak from the components, but a system of small loop flux and node charge variables that are normally neglected by Kirchhoff's laws. The high frequency oscillations arise as the small electric and magnetic fields within the circuit loops change and interact with one another and with the large low frequency oscillations within the components of the circuit.  They are a fundamental source of noise that arises from the geometry of the circuit itself.

\end{document}